\documentclass[12pt]{article}
\usepackage{epsfig,epsf}
\usepackage{axodraw}
\usepackage{amsmath,amssymb}
\usepackage{paralist}
\usepackage{slashed}
\usepackage{graphicx}
\usepackage{subfig}
\usepackage{color}
\usepackage{fullpage}
\usepackage{multirow}
\usepackage{hyperref}
\usepackage{mcite}

\def\lsim{\raise0.3ex\hbox{$\;<$\kern-0.75em\raise-1.1ex\hbox{$\sim\;$}}}
\def\gsim{\raise0.3ex\hbox{$\;>$\kern-0.75em\raise-1.1ex\hbox{$\sim\;$}}}

\newcommand{\beq}{\begin{equation}}
\newcommand{\eeq}{\end{equation}}

\newcommand{\bmat}{\left(\begin{array}}
\newcommand{\emat}{\end{array}\right)}
\newcommand{\be}{\begin{equation}}
\newcommand{\ee}{\end{equation}}
\newcommand{\bea}{\begin{eqnarray}}
\newcommand{\eea}{\end{eqnarray}}

\usepackage{graphics}
\usepackage{epsfig}
\usepackage{pict2e}
\begin{document}
\begin{flushright}
SHEP-12-29\\
\today
\end{flushright}
\vspace*{1.0truecm}
\centerline{\bf\Large Variations on a Higgs theme}
\vspace*{0.75truecm}
\centerline{\large S. Moretti}
\vspace*{0.5truecm}
\centerline{\sl School of Physics and Astronomy, University of 
Southampton,}
\vspace*{0.25truecm} 
\centerline{\sl Highfield, Southampton SO17 1BJ, UK.}
\vspace*{1.0truecm}
\begin{abstract}
\noindent We show how the $Z$ boson can be generated in gluon-gluon fusion and yield two photons, via $gg\to Z\to \gamma\gamma$, through massive fermion loops only,
thereby contributing events to a candidate Higgs sample in the di-photon channel. A sub-leading contribution also exists from $q\bar q\to Z\to\gamma\gamma$ events.  
Assuming the Standard Model, the corresponding event rates are negligible at the LHC stages
of 7, 8 TeV, given the luminosities collected therein (about 5 and 20 fb$^{-1}$, respectively). Conversely, at 14 TeV, the first process become accessible for luminosities 
of order 300 fb$^{-1}$. Finally, we show how additional fermion states entering such loops, in production, in decay or in both cases,
could affect the predictions in this channel by curiously mimicking Higgs signals. 
\end{abstract}
\newpage
\section{Introduction}
About a couple of years ago, the ATLAS and CMS experimental collaborations at the
Large Hadron Collider (LHC) at CERN had announced the observation of a
new boson, with a mass of about 125 GeV \cite{Aad:2012tfa,Chatrchyan:2012ufa},
which was consistent with a Higgs particle, $H$, the last undiscovered
object in the Standard Model (SM)\footnote{Some supplemental evidence was also gathered by CDF and 
D0 at FNAL \cite{HiggsEvidenceTevatron1,HiggsEvidenceTevatron2}.}. The most recent results reported by ATLAS~\cite{ATLASNOTE2:2013,ATLAS-CONF-2013-072} and 
CMS~\cite{:2013tq, Chasco:2013pwa, CMS-PAS-HIG-13-005, CERN-PH-EP-2014-117} confirm such a
Higgs boson discovery beyond doubt.  

The decay channels investigated experimentally with highest
precision are
$H \to \gamma \gamma $, $H\to ZZ \to 4 l$ and $H \to WW \to l \nu_l l^{(')} \nu_{l^{(')}}$, where $l^{(')}$ denotes a lepton
(electron and/or muon) and $\nu$ its associated neutrino. 
While these data are indeed compatible with the SM, they also indicate presently a small enhancement in the di-photon
mode, with respect to SM. Indeed, the di-photon sample can be very sensitive to possible Beyond the SM (BSM) effects, owing to the fact that, amongst the aforementioned 
SM-like decay modes
(or else the $b\bar b$ and $\tau^+\tau^-$ ones, to which ATLAS and CMS also have sensitivity, albeit reduced in comparison \cite{HtoFermions}), it is the only one in which such effects can enter at the same perturbative order as the SM ones, in the
triangle loop connecting the Higgs boson to the two photons\footnote{A similar phenomenology also occurs in the $Z\gamma$ case, however, experimental sensitivity to this channel is much smaller with respect to the $\gamma\gamma$ case,
so that only upper limits exist to date in this mode.}. 

Fuelled by the fact that some analyses (at time with partial luminosity only) of
di-photon data have seen over the past couple of years significant deviations from the SM predictions, a flurry of literature emerged trying to explain the latter in one or another BSM scenario. Far from endorsing either a SM or BSM hypothesis, we 
simply like to study here a forgotten contribution to the di-photon background which arises entirely in the SM. This is induced by $gg\to Z
\to \gamma\gamma$,
where the $Z$ boson is never on mass-shell.
In fact, contrary to a 
rather widespread popular belief in the community, following the detection of the $H\to\gamma\gamma$ decay mode,
a spin-1 state can produce two photons, as the Landau-Yang theorem \cite{LandauYang}, often erroneously invoked, is actually only 
applicable to on-shell particles \cite{Zhemchugov,Kanda}.  (Similar arguments apply to the case of $gg\to Z$.) For the case 
of the $Z$ boson of the SM, the coupling is induced in both $gg\to Z$ ~`production' and $Z\to\gamma\gamma$ ~`decay' solely via a triangle loop of heavy quarks 
(primarily the top one then) through the Goldstone component of the $Z$ propagator (hence the effect is best seen in
non-unitary gauges), which is in fact pseudoscalar in its couplings, hence in turn explaining why the $W^+W^-$ loop does 
not enter the $Z$ di-photon transition. We will compare the yield of the $gg\to Z\to \gamma\gamma$
and (also) $q\bar q\to Z\to \gamma\gamma$ processes against 
that of the SM Higgs process $gg\to H\to \gamma\gamma$ (occurring via two triangle loops, of quarks only in production and both bosons and fermions in decay) as well those of the 
customary backgrounds $gg\to \gamma\gamma$ (occurring via a box loop of quarks) and $q\bar q\to \gamma\gamma$ (occurring at tree level).

The plan of the paper is as follows. In the next section we describe the calculation. In the following one we discuss
our results. We then conclude. We also have an appendix containing some key formulae.   

\section{Calculation}

Contrary to the claim made by hundreds of papers\footnote{That we do not intend to quote here for obvious reasons, though a knowledgeable reader will be able to source these.
We cite instead this paper \cite{Ralston:2012ye}, which also highlighted that this statement was wrong, though the solution proposed
therein is different from ours, as the author illustrates possibly {\it resonant} $\gamma\gamma$ production through
interference effects between a scalar and vector propagator, with suitable relative complex phase. } stating that, if an excess is seen in the $\gamma\gamma$ decay channel, this cannot be produced by a spin-1 particle, we show here how this is possible. The very same papers 
often misleadingly quote the Landau-Yang's theorem \cite{LandauYang}, in order to support their claim.
The latter, as originally formulated, is however only applicable to on-shell objects 
\cite{Zhemchugov,Kanda} and is instead
violated when, e.g., the $\gamma\gamma$ pair is produced by a $Z$ boson which is not on-shell
(i.e., non-resonant). Similar arguments can be applied to off-shell $Z$ production via $gg$ fusion.
Therefore, the existence of the processes $gg\to Z\to \gamma\gamma$ 
and $q\bar q\to Z\to \gamma\gamma$ is perfectly legitimate. However, they can only occur
through the pseudoscalar component of the (off-shell) $Z$ boson: in other words, according to the Becchi-Stora-Rouet-Tyutin (BRST)
\cite{BRST}
equalities, through its associated Goldstone component (in a generic $R_\xi$ 
gauge)\footnote{Incidentally,
the Goldstone of the $Z$ originates in the Higgs doublet, hence it is not surprising that the $Z$ boson can produce $\gamma\gamma$ pairs,
just like the Higgs boson does.}. This is made evident if one uses the Landau gauge ($\xi\to0$) for the calculation  of the
above process. The very fact that it is the pseudoscalar component of the $Z$ boson to appear in it also means that 
in the $Z\to\gamma\gamma$ transition only fermion loops are involved, not (charged) gauge boson ones. 

Two different calculations have been performed, so as to cross check one another. Furthermore, one computation was done 
analytically and the other numerically. In particular, the latter
 was done in the Landau gauge while the former in the unitary gauge. So, we are bound to use the unitary gauge ($\xi\to\infty$) to
illustrate the calculation, which makes it more cumbersome yet more physically intuitive. But before doing so, let us list the inputs required to perform
our numerical computations.

The relevant numerical inputs adopted here were as follows. For the top mass and width
 we have taken $m_t=175$ GeV and $\Gamma_t=1.55$ GeV, respectively. 
The $Z$ mass used was $M_Z=91.19$ GeV and was related to the $W$ mass, $M_W$, via the
SM formula $M_W=M_Z\cos\theta_W$, where $\sin^2\theta_W=0.232$.
(Corresponding widths were $\Gamma_Z=2.5$ GeV and $\Gamma_W=2.08$ GeV.)
For the Higgs boson we have taken $M_H=125$ GeV and $\Gamma_H=4.2$ MeV. For the Electro-Magnetic (EM) coupling constant
we have taken $\alpha=1/128$ throughout.
The Parton Distribution Functions (PDFs) that we have used are the CTEQ5L set
\cite{cteq5} taken at the factorisation/renormalisation scale $Q=\mu=\sqrt{\hat s}$. (We also have checked
other PDFs and adopted different scale choices, but found no significant difference in the relative size of the
processes studied here.) The choice of PDFs dictates the running and parameters used
to compute $\alpha_s$. All rates are presented at the LHC energies of 7, 8  and 14 TeV\footnote{As we are only interested in the relative strenght of the aforementioned di-photon
process, we do not include any strong or EW corrections in our analysis.}. 
 
\subsection{The computation of $gg\to Z\to \gamma\gamma$}

\noindent {We look first at the $Z\to \gamma\gamma$ transition and start with the following definitions:} 
\begin{itemize}
\item $p_1^\mu, \ p_2^\mu$ are the momenta of the two outgoing photons;
\item $\epsilon_1^\mu, \ \epsilon_2^\mu$   are the polarisation vectors of the two outgoing photons;
\item $q^\mu \,(=p_1^\mu+p_2^\mu)$ is the momentum of the incoming $Z$;
\item $\epsilon_z^\mu$ is the polarization vector of the $Z$.
\end{itemize}
The photons are on-shell and so we have
\begin{equation} p_1\cdot\epsilon_1  \ =  \  p_2\cdot \epsilon_2 \ =  0, \end{equation} 
but the $Z$ is off-shell and so we have
\begin{equation}  q\cdot \epsilon_z \ \neq \ 0. \end{equation} 
We may in general expand the polarization vector of the $Z$ as
\beq 
 \epsilon_z^\mu \ = \ \frac{\epsilon_z \cdot q}{q^2} q^\mu -\epsilon_z\cdot \epsilon_1^* \epsilon_1^\mu
-\epsilon_z\cdot \epsilon_2^* \epsilon_2^\mu - 
\frac{\epsilon_z \cdot (p_1-p_2)}{q^2} \left(p_1^\mu-p_2^\mu \right). 
 \label{ez} \eeq
The relation in unitary gauge
\beq \sum_{\lambda} \epsilon_z^\mu(\lambda) \epsilon^\nu(\lambda)  \ = \ -g^{\mu\nu} + \frac{q^\mu q^\nu}{M_Z^2} \eeq
leads to 
\beq \epsilon_z \cdot q \ = \ \sqrt{\frac{q^2}{M_Z^2} (q^2-M_Z^2)}. \label{ezq} \eeq
This vanishes when the $Z$ goes on shell (as expected).

If we allow the matrix element for the vertex $Z \to \gamma \gamma$
to be a vector with index $\mu$ we do not need to discuss the polarisation
vector of the $Z$. We show that this vertex is proportional to $q^\mu$ where
 $q$ is the momentum of the $Z$.

\bigskip

The amplitude (wherein the helical lines represent either photons or gluons whereas the wavy line refers to the $Z$) 
\vspace*{0.25cm}\noindent
\begin{center}
\begin{picture}(200,120)
\Photon(10,60)(60,60){4}{4}
\Gluon(110,10)(160,10){4}{4}
\Gluon(110,110)(160,110){4}{4}
\ArrowLine(60,60)(110,10) \ArrowLine(110,10)(110,110) \ArrowLine(110,110)(60,60)
\put(0,55){$\mu$} \put(165,5){$\epsilon_1$}\put(165,105){$\epsilon_2$}
 \put(130,19){$p_1$} \put(130,119){$p_2$}  \put(20,69){$q$}
 \put(113,55){$k$}
\end{picture}
\end{center}
\vspace*{0.25cm}\noindent
is not gauge invariant for an off-shell $Z$ and we work in the gauge
\beq p_2\cdot\epsilon_1  \ =  \  p_1\cdot \epsilon_2 \ =  0. \eeq
\bigskip

Such an amplitude is given by (wherein $m$ is the fermion mass)
 \beq {\cal A} \  = \ \ e^2 C_A \int \frac{d^4k}{(2\pi)^4}
 \frac{\left({\cal N}_1 +{\cal N}_2\right)}{(k^2-m^2)((k-p_2)^2-m^2)((k+p_1)^2-m^2)}
 \eeq
where
\beq{\cal N}_1 \ = \ \mathrm{Tr} \left(\gamma \cdot \epsilon_1 (\gamma \cdot k+m)
 (\gamma \cdot (k-p_2)+m)  \gamma^\mu  \left(g_V+g_A \gamma^5\right)
     (\gamma \cdot (k+p_1)+m) \right) \eeq
from the graph shown
and
\beq {\cal N}_2 \ = \ -\mathrm{Tr} \left(\gamma \cdot \epsilon_1 (\gamma \cdot (k+p_1)-m)
 \gamma^\mu \left(g_V+g_A \gamma^5\right)
 (\gamma \cdot (k-p_2)-m) \gamma \cdot \epsilon_2
     (\gamma \cdot k-m) \right) \eeq 
from the graph with the fermions circulating in the reverse direction.
The factor of $C_A(=3)$ comes from summing over all colours of internal quarks. 

These terms are written in terms of the Passarino-Veltman \cite{PV} functions defined in the usual way.
In the case of equal masses we have the following relations:
\begin{enumerate}
\item[a.] $$ B_1(q^2,m^2,m^2) \ = \ -\frac{1}{2} B_0(q^2,m^2,m^2) $$
\item[b.] $$ C_{11}(0,q^2,0,m^2,m^2,m^2) \ = \ \frac{2}{q^2} B_0(q^2,m^2,m^2) $$
\item[c.] $$ C_{12}(0,q^2,0,m^2,m^2,m^2) \ = \ \frac{1}{q^2} B_0(q^2,m^2,m^2) $$
\item[d.] $$ 4 C_{24}(0,q^2,0,m^2,m^2,m^2) \ = \ 1 + B_0(q^2,m^2,m^2)  +2 m^2   C_{0}(0,q^2,0,m^2,m^2,m^2) $$
\item[e.]  $$ C_{21} (0,q^2,0,m^2,m^2,m^2) \ = \    \frac{1}{q^2} \left(1 - B_0(q^2,m^2,m^2)
 +2 m^2   C_{0}(0,q^2,0,m^2,m^2,m^2)  \right)$$
\item[f.]  $$ C_{22} (0,q^2,0,m^2,m^2,m^2) \ = \    -\frac{1}{2q^2}  B_0(q^2,m^2,m^2) $$
\item[g.] $$ C_{23} (0,q^2,0,m^2,m^2,m^2) \ = \    \frac{1}{2q^2} \left(1 - B_0(q^2,m^2,m^2)
 +2 m^2   C_{0}(0,q^2,0,m^2,m^2,m^2)  \right)$$
\end{enumerate}
so that everything can be expressed in terms of two master integrals:
\beq  B_0(q^2,m^2,m^2) \ =  \ -\int_0^1 d\alpha \ln\left(1-\frac{s\alpha(1-\alpha)}{m^2} \right) \eeq
and
\beq  C_0(0,q^2,0,m^2,m^2,m^2) \ =  \ \int_0^1 \frac{d\alpha}{\alpha} \ln\left(1-\frac{s\alpha(1-\alpha)}{m^2} \right).\eeq
These two integrals can in turn be expressed in terms of logarithms and di-logarithms, as seen in the appendix.

Taking the trace, applying the above relations  and developing the kinematics, we find (after some
manipulations carried out using FORM \cite{FORM}) the following relation:

\beq{\cal A} \ = \ \frac{\alpha}{\pi} g_A C_A
 \left[
     \frac{q^\mu}{q^2} 
       \epsilon_{\sigma\nu\rho\tau}p_1^\sigma p_2^\nu \epsilon_1^\rho \epsilon_2^\tau
             \left(B_0-1-2m^2C_0\right) + 
      \frac{1}{2}
        \epsilon_{\sigma\nu\rho}^{ \ \ \ \mu}(p_1- p_2)^\sigma \epsilon_1^\nu \epsilon_2^\rho 
             \left(1-B_0-2m^2C_0\right)
 \right]. \eeq

We remark here that, as expected, the amplitude is proportional to $g_A$, the axial coupling of he $Z$
to fermions. In fact, for the vector coupling, the amplitude vanishes identically by Furry's theorem \cite{Furry}.

Now we note that
\beq        \epsilon_{\sigma\nu\rho\tau}p_1^\sigma p_2^\nu \epsilon_1^\rho \epsilon_2^\tau
 \ = \ \frac{1}{2}
 \epsilon_{\sigma\nu\rho\tau}(p_1-p_2)^\sigma(p_1+ p_2)^\nu \epsilon_1^\rho \epsilon_2^\tau
\eeq

and, since $p_1,p_2,\epsilon_1,\epsilon_2$ are mutually orthogonal ($s=(p_1+p_2)^2$),

\beq   \epsilon_{\sigma\nu\rho}^{ \ \ \ \mu}(p_1- p_2)^\sigma \epsilon_1^\nu \epsilon_2^\rho 
 \ = \  \frac{q^\mu}{s}  \epsilon_{\sigma\nu\rho\tau}(p_1-p_2)^\sigma(p_1+ p_2)^\nu \epsilon_1^\rho \epsilon_2^\tau.
\eeq

This amplitude is thus reduced to

\beq
{\cal A} \ = \ -2  \frac{q^\mu}{s} \frac{\alpha}{\pi} g_A m^2  C_0   C_A  
 \epsilon_{\sigma\nu\rho\tau}(p_1-p_2)^\sigma(p_1+ p_2)^\nu \epsilon_1^\rho \epsilon_2^\tau. 
\eeq

We  also have a similar term from the production vertex from gluon-gluon scattering
with $\alpha$ replaced by $\alpha_s$ and the colour factor $C_A$ replaced by
 $\frac{1}{2} \delta_{ab} $  ($a,b$ are the gluon colours).

The full  amplitude for the process
$$ g(p_1,\lambda_1)+ g(p_2,\lambda_2) \ \to \ Z \ \to \   \gamma(p_3,\lambda_3)+ \gamma(p_4,\lambda_4)
$$
is therefore given by

\begin{eqnarray}
{\cal A} & = &  4 C_A \delta_{ab} \frac{\alpha \alpha_s}{ s^2\pi^2}
  \sum_{i,j} g_A^{(i)}g_A^{(j)} \,  m_i^2 m_j^2  \,
C_0(0,s,0,m_i^2,m_i^2,m_i^2) \, C_0(0,s,0,m_j^2,m_j^2,m_j^2) \,
 \nonumber \\ & & \ \times \  
 \left[    \epsilon_{\sigma\nu\rho\tau}(p_1-p_2)^\sigma(p_1+ p_2)^\nu \epsilon_1^\rho \epsilon_2^\tau
 \right] \left[ 
  \epsilon_{\sigma^\prime\nu^\prime\rho^\prime\tau^\prime}
(p_3-p_4)^{\sigma^\prime}(p_3+ p_4)^{\nu^\prime} \epsilon_3^{\rho^\prime} \epsilon_4^{\tau^\prime}
 \right]   \ \nonumber \\
   & & 
\ \times q^\alpha q^\beta \left(\frac{g_{\alpha\beta} - \frac{q_\alpha q_\beta}{M_Z^2}}{(s-M_Z^2)}
 \right)
 \nonumber  \end{eqnarray}
where the $\sum_{i,j}$ goes over all flavours of internal quarks and is dominated by the 
top one (the contribution from lepton loops in the $Z$ decay being negligible).

Now,
\beq q^\alpha q^\beta \left(g_{\alpha\beta} - \frac{q_\alpha q_\beta}{M_Z^2} \right)
 \ = \ s-\frac{s^2}{M_Z^2}, \eeq
so we see that the $Z$-pole cancels and we get

\begin{eqnarray}
{\cal A} & = &  4 C_A \delta_{ab} \frac{\alpha \alpha_s}{M_Z^2 s\pi^2}
  \sum_{i,j} g_A^{(i)}g_A^{(j)} \,  m_i^2 m_j^2  \,
C_0(0,s,0,m_i^2,m_i^2,m_i^2) \, C_0(0,s,0,m_j^2,m_j^2,m_j^2) \,
 \nonumber \\ & & \ \times \  
 \left[    \epsilon_{\sigma\nu\rho\tau}(p_1-p_2)^\sigma(p_1+ p_2)^\nu \epsilon_1^\rho \epsilon_2^\tau
 \right] \left[ 
  \epsilon_{\sigma^\prime\nu^\prime\rho^\prime\tau^\prime}
(p_3-p_4)^{\sigma^\prime}(p_3+ p_4)^{\nu^\prime} \epsilon_3^{\rho^\prime} \epsilon_4^{\tau^\prime}
 \right].  
 \nonumber  \end{eqnarray}

\subsection{The computation of $q\bar q\to Z\to \gamma\gamma$}

For the process
$$ q \, \bar{q} \ \to \ Z \ \to \ \gamma \, \gamma $$
the vertex representing gluon-gluon fusion is replaced by the coupling of the $Z$ to the incoming quarks,
\beq \overline{v}(\lambda_1,p_1) \gamma^\mu \left(g_v-g_A \gamma^5 \right) u(\lambda_2,p_2). \eeq
The vertex from the $Z$ decay still projects out the polarisation proportional
to $q^\mu$ and so we may rewrite this (after the projection) as ($m$ is the quark mass)
 \beq \frac{q^\mu q^\nu}{s} \overline{v}(\lambda_1,p_1) \gamma^\nu (g_v-g_A \gamma^5) u(\lambda_2,p_2) 
 \ = \ \frac{q^\mu}{s} g_A 2m  \overline{v}(\lambda_1,p_1) \gamma^5 u(\lambda_2,p_2) 
 \ = \ \frac{q^\mu}{\sqrt{s}} g_A 2m  \delta_{\lambda_1,-\lambda_2} \eeq
using
 \beq \overline{v}(\lambda_1,p_1)\gamma^5 u(\lambda_2,p_2)
  \ = \ \sqrt{2 p_1\cdot p_2}  \delta_{\lambda_1,-\lambda_2}, \eeq 
for massless fermions.

Thus the amplitude from the process with  incoming quark-antiquark pairs of mass $m_i$
and colours $i,j$ is
\beq
{\cal A} \ =  \  4 C_A \delta_{ij}  \delta_{\lambda_1,-\lambda_2}
      \frac{\alpha}{M_Z^2 \sqrt{s} \pi} m_i g_A^{(i)}
  \sum_{j} g_A^{(j)} \,   m_j^2  \, C_0(0,s,0,m_j^2,m_j^2,m_j^2) 
  \epsilon_{\sigma^\prime\nu^\prime\rho^\prime\tau^\prime}
(p_3-p_4)^{\sigma^\prime}(p_3+ p_4)^{\nu^\prime} \epsilon_3^{\rho^\prime} \epsilon_4^{\tau^\prime}. 
\eeq

Furthermore, we have the relations
\beq \epsilon_{\sigma^\prime\nu^\prime\rho^\prime\tau^\prime}
(p_3-p_4)^{\sigma^\prime}(p_3+ p_4)^{\nu^\prime} \epsilon_3^{\rho^\prime} \epsilon_4^{\tau^\prime}
 \ = \ s\delta_{\lambda_3,\lambda_4} \eeq
and
\beq
   m_j^2  \, C_0(0,s,0,m_j^2,m_j^2,m_j^2)  \ = \  -2 f\left(\frac{4m_j^2}{s}\right) \eeq
with $f(\tau)$ as in, e.g., Ref.~\cite{HHG}.

Therefore
\beq
{\cal A} \ =  \  -8 C_A \delta_{ij}  \delta_{\lambda_1,-\lambda_2}
      \frac{\alpha \sqrt{s}}{M_Z^2 \pi} m_i g_A^{(i)}
  \sum_{j} g_A^{(j)} \,  f\left(\frac{4m_j^2}{s}\right)  \,  \delta_{\lambda_3,\lambda_4}. \eeq

Also (here $g_W=e/\sin\theta_W$),
\beq \frac{g_A}{M_Z} \ = \ \pm \frac{1}{2} \frac{g_W}{M_W} \eeq 
and so we have
finally

\beq
{\cal A} \ =  \  \pm 2 \frac{g_W^2}{M_W^2} C_A \delta_{ij}  \delta_{\lambda_1,-\lambda_2}
      \frac{\alpha \sqrt{s}}{ \pi} m_i 
  \sum_{j} (-1)^j \,  f\left(\frac{4m_j^2}{s}\right)  \,  \delta_{\lambda_3,\lambda_4}.
\eeq

Note the $\delta$-functions for {\it both} the incoming quark helicities (which must be opposite
 since the mass insertion requires helicity flip) and the outgoing photons (which must have the
same helicity to conserve angular momentum).

\section{Results}
The total (inclusive) cross sections (in fb) for the processes $gg\to Z\to \gamma\gamma$ and
$q\bar q \to Z\to \gamma\gamma$ at the LHC with $\sqrt s=7,8$ and 14 TeV is found in
Tab. \ref{tab:Xsects}. Given the accumulated luminosities at the first two stages of the LHC,
5 (at 7 TeV) and 20 (at 8 TeV) fb$^{-1}$, it is clear that neither of the processes is accessible therein. In contrast,
at the highest energy stage (i.e., 14 TeV) with design luminosity (say, 300 fb$^{-1}$), one should expect
some 17 events from $gg$ fusion and again none from $q\bar q$ scatterings, at inclusive
level. If the typical SM Higgs selection cuts
\beq
p^T_\gamma    >  20~{\rm GeV},\quad
|\eta_\gamma|    <  2.5, \quad
M_{\gamma\gamma}    >  100~{\rm GeV},
\label{cuts}
\eeq
in transverse momentum ($p^T$), pseudorapidity       ($\eta$) and invariant mass
($M_{\gamma\gamma}$) of the photons, are enforced, then the detectable events (at $\sqrt s=14$ TeV with
${\cal L} dt=300$ fb$^{-1}$ for the $gg$ case) scale down to 10.

These are rather small numbers and, if regarded as contributors to a candidate 
Higgs sample at the LHC (again, with design energy and luminosity), they are very subleading with
respect to both the $gg\to H\to \gamma\gamma$ signal (yielding
47 fb after the cuts in eq. (\ref{cuts})
and the other known background in the $gg$ channel, i.e.,  
$gg\to Box\to \gamma\gamma$ (giving 793 fb after cuts). In fact, it should be noted that
the dominant di-photon background is the tree-level $q\bar q\to \gamma\gamma$, as it produces
6770 fb of cross section (after cuts). In the light of these results, we will then
neglect from now on discussion of the $q\bar q\to Z\to \gamma\gamma$ process, apart from
a reference histogram in the upcoming figure\footnote{We
 have also investigated possible interference effects between the
$q\bar q\to Z\to \gamma\gamma$ amplitude and the $q\bar q\to \gamma\gamma$ one
and found them negligible.}.

\begin{figure}[!t]
\centering
\includegraphics[width=0.5\linewidth,angle=90]{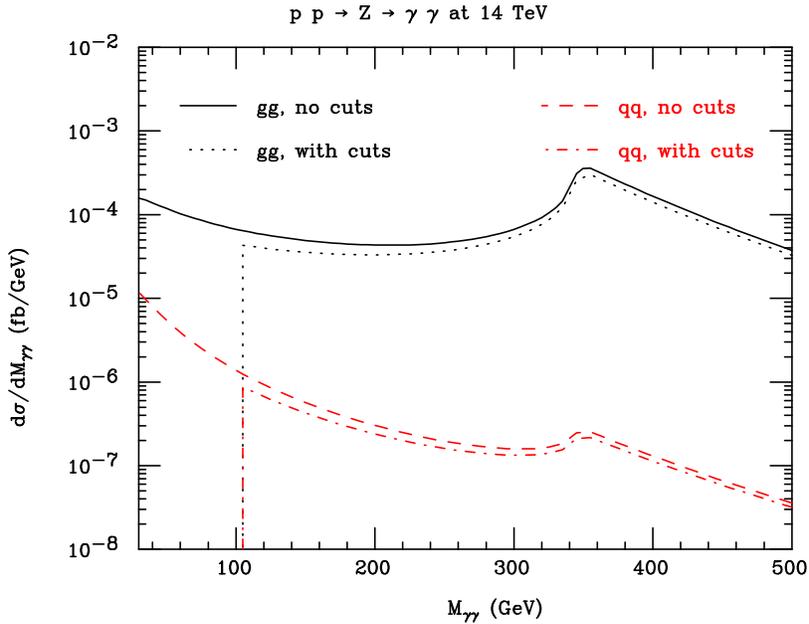}
\caption{Differential di-photon mass distributions at the 14 TeV LHC for the processes $gg\to Z\to \gamma\gamma$ and
$q\bar q\to Z\to \gamma\gamma$ before and after the cuts 
$p^T_\gamma>20$ GeV, $|\eta_\gamma|<2.5$ and $M_{\gamma\gamma}>100$ GeV.
CTEQ(5L) with $Q=\mu=\sqrt{\hat s}$ is used.}
\label{fig:ggZ+qqZ-mass}
\end{figure}

The differential distributions in the di-photon invariant mass for the two processes
under consideration, $gg\to Z\to \gamma\gamma$ and $q\bar q\to Z\to \gamma\gamma$, at
14 TeV, are found in Fig. ~\ref{fig:ggZ+qqZ-mass}. This shows that much of the 
cross section is located around the top-antitop threshold, $M_{\gamma\gamma}\approx 2m_t$,
for both channels. This confirms, as expected, the dominance of the top contribution in the 
triangle loops. Furthermore, notice that the increase at threshold is more pronounced for the
$gg$ subchannel, in comparison to the $q\bar q$ one, owing to the fact that the 
aforementioned loop appears both in production and decay for $gg\to Z\to \gamma\gamma$ whereas only in decay for $q\bar q\to Z\to \gamma\gamma$.  This pattern is the same before and after the cuts in eq. (\ref{cuts}). 

\begin{figure}[!t]
\centering
\includegraphics[width=0.5\linewidth,angle=90]{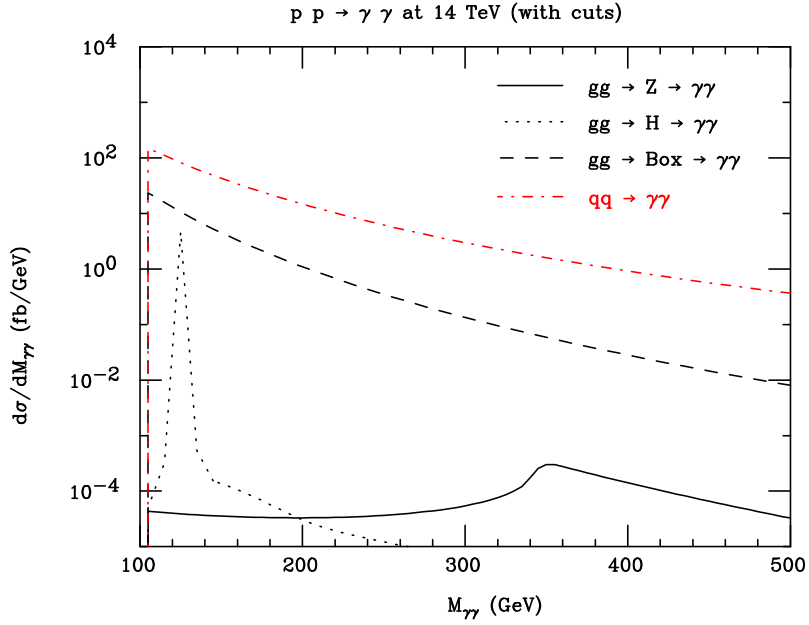}
\caption{Differential di-photon mass distributions at the 14 TeV LHC for the processes 
$gg\to Z\to \gamma\gamma$,
$gg\to H\to \gamma\gamma$,
$gg\to Box\to \gamma\gamma$ 
and
$q\bar q\to \gamma\gamma$ after the cuts
$p^T_\gamma>20$ GeV, $|\eta_\gamma|<2.5$ and $M_{\gamma\gamma}>100$ GeV.
CTEQ(5L) with $Q=\mu=\sqrt{\hat s}$ is used.}
\label{fig:all-mass}
\end{figure}

Fig. \ref{fig:all-mass} illustrates again the subdominance of the $gg\to Z\to \gamma\gamma$
process with respect to the others mentioned above, i.e.,
$gg\to H\to \gamma\gamma$,
$gg\to Box\to \gamma\gamma$
and
$q\bar q\to \gamma\gamma$, now seen in the di-photon invariant mass. This is shown 
after the aforementioned cuts. Yet, with
increasing di-photon invariant masses, up to around $2m_t$ and onwards, the relative importance
of $gg\to Z\to \gamma\gamma$ with respect to the other channels grows steadily,
reaching in such an invariant mass region the $0.2$ permille level with respect to the leading 
$q\bar q\to \gamma\gamma$ term. (Notice that the curves are obtained from a fit to histograms which are 10 GeV wide, hence the distorted shape of the (otherwise very 
narrow) $H$ peak.)

\begin{figure}[!t]
\centering
\includegraphics[width=0.5\linewidth,angle=90]{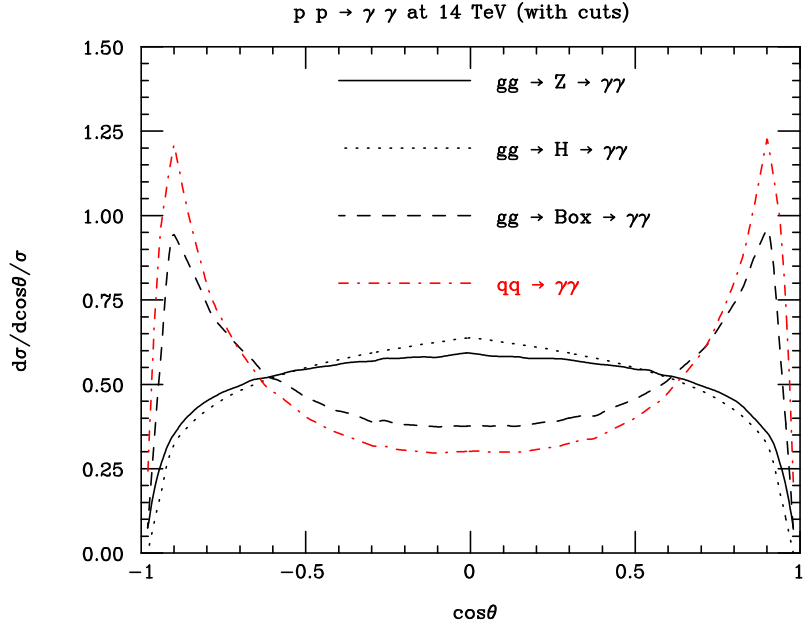}
\caption{Differential polar angle (defined in the CM rest frame) distributions at the 14 TeV LHC for the processes
$gg\to Z\to \gamma\gamma$,
$gg\to H\to \gamma\gamma$,
$gg\to Box\to \gamma\gamma$
and
$q\bar q\to \gamma\gamma$ after the cuts
$p^T_\gamma>20$ GeV, $|\eta_\gamma|<2.5$ and $M_{\gamma\gamma}>100$ GeV.
CTEQ(5L) with $Q=\mu=\sqrt{\hat s}$ is used. Note the normalisation to unity.}
\label{fig:all-cost}
\end{figure}

The $gg\to Z\to \gamma\gamma$ channel becomes relatively more important, with respect to the other di-photon
backgrounds, if viewed differentially, in the polar angle of either of the photons, $\theta$, 
when defined in the rest frame
of the Center-of-Mass (CM). This observable, as it is well known, is sensitive to both the spin and CP-properties
of the Higgs boson and it has been extensively used for this purpose by the ATLAS and CMS collaborations
(see, e.g., Ref. \cite{SpinCP}). By looking at Fig. \ref{fig:all-cost}, it is clear that, while the 
$gg\to Box\to \gamma\gamma$
and
$q\bar q\to \gamma\gamma$ backgrounds have a completely different structure in $\cos\theta$ with respect to
the $gg\to H\to \gamma\gamma$ signal, the shape of $gg\to Z\to \gamma\gamma$ is very similar to it.
Hence, in this observable more than others, one ought to achieve an accurate modelling of the complete background, including
the contribution from the $gg$ process that we have computed here. 

\begin{figure}[!t]
\centering
\includegraphics[width=0.5\linewidth,angle=90]{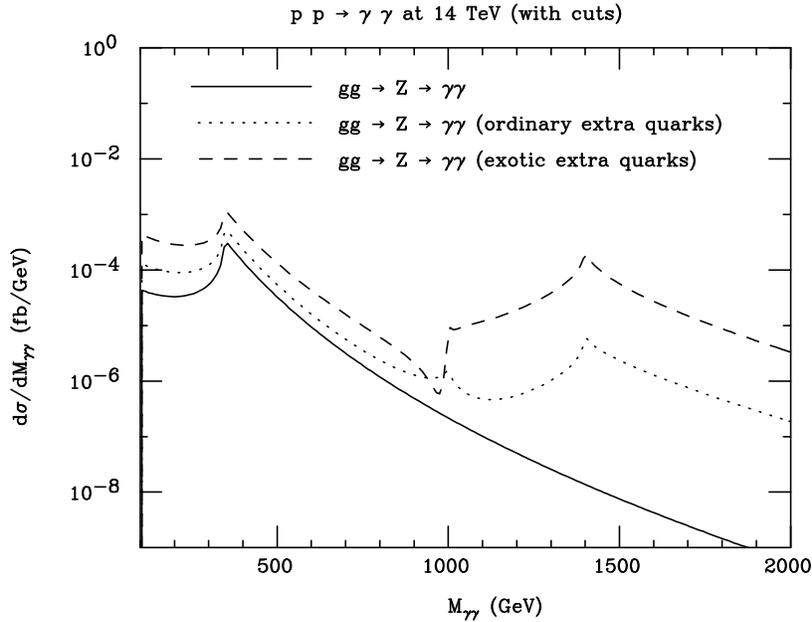}
\caption{Differential di-photon mass distributions at the 14 TeV LHC for the processes
$gg\to Z\to \gamma\gamma$,
$gg\to Z\to \gamma\gamma$ including one additional generation of ordinary quarks
$gg\to Z\to \gamma\gamma$ including one additional generation of exotic quarks
after the cuts
$p^T_\gamma>20$ GeV, $|\eta_\gamma|<2.5$ and $M_{\gamma\gamma}>100$ GeV.
CTEQ(5L) with $Q=\mu=\sqrt{\hat s}$ is used.}
\label{fig:ggZaa-newmass}
\end{figure}

Before closing, we would also like to emphasise that the $gg\to Z\to \gamma\gamma$ process, in virtue
of its loop structure that is doubly sensitive to virtual heavy (coloured) fermions (i.e., both in production
and decay), can in principle reveal the presence of additional states of this kind, with respect to the SM, which may or may not be accessible 
via direct searches. Fig. \ref{fig:ggZaa-newmass} illustrates this for the case of, e.g., 
one additional generation of up- and -down-type vector-like quarks\footnote{Notice that the presence 
of a possible fourth generation of SM-type fermions is less and less favourite by the 
LHC Higgs data, see  \cite{fourth} for a review.}, with both standard ($+2/3$ and $-1/3$, respectively) and exotic ($+5/3$ and $-4/3$, respectively)
EM charges, both cases with masses 700 (up-type) and 500 (down-type) GeV. (These 
states are predicted by various theoretical frameworks like, e.g., 
composite Higgs models\footnote{Our chosen illustrative case is from the so-called 4-Dimensional Composite Higgs
Model (4DCHM) of Ref. \cite{4DCHM}, as implemented in \cite{4DCHMHiggsPaper}, wherein the above
masses are consistent with experiment \cite{VLQLimits}.}, little Higgs models, scenarios 
with extra dimensions, models with gauging of the flavour group,
non-minimal supersymmetric  extensions as well as 
Grand Unified Theories.) As expected, we notice in the plotted spectra the additional thresholds at (twice)
the above two masses\footnote{Despite the fact that one cannot and will  not be able to reconstruct the Breit-Wigner
shaped `Higgs resonance' from data, owing to the fact that the Higgs boson of the SM is 4.2 MeV wide at 125 GeV and 
the di-photon mass resolution of detectors is and will remain far larger than this, so that data actually
shape a Gaussian `Higgs enhancement', we find it of  
 very speculative value and of little phenomenological relevance, at least to our judgment,
to mention here that a 62.5 GeV fermion could actually easily conspire to produce a 125 GeV `threshold enhancement', 
very nearly Gaussian in shape.}.

\begin{table}[!h]
\captionsetup[subfloat]{labelformat=empty,position=top}
\centering
\subfloat[$gg\to Z\to \gamma\gamma$]
{
\begin{tabular}{|l|l|l|l|}
\hline
 & 7 TeV & 8 TeV & 14 TeV\\
\hline
Before cuts&$0.014$&$0.018$&$0.055$\\
After cuts&$0.0067$&$0.0095$&$0.034$\\
\hline
\end{tabular}
}
\subfloat[$q\bar q\to Z\to \gamma\gamma$]
{\begin{tabular}{|l|l|l|l|}
\hline
 & 7 TeV & 8 TeV & 14 TeV\\
\hline
Before cuts&$0.00045$&$0.00053$&$0.0010$\\
After cuts&$2.21\times10^{-5}$&$2.95\times10^{-5}$&$9.02\times10^{-5}$\\
\hline
\end{tabular}
}
\caption{\label{Tab:sigma:M}
Cross section in fb for $gg\to Z\to \gamma\gamma$ 
and $q\bar q\to Z\to \gamma\gamma$ 
at the three  LHC energy stages. The selection enforced employs
the following cuts: 
$ p^T_\gamma    >  20$    GeV,
$|\eta_\gamma|    <$  2.5 and
$M_{\gamma\gamma}    >  100$ GeV.
CTEQ(5L) with $Q=\mu=\sqrt{\hat s}$ is used.}
\end{table}
\label{tab:Xsects}

\section{Conclusions}
We have studied the yield of the two processes $gg\to Z\to \gamma\gamma$ and 
$q\bar q\to Z\to \gamma\gamma$ in the SM at the LHC.
These two channels emerge only when the intermediate $Z$ boson is off-shell, so that they are never large,
though the $gg$ subchannel can be accessed at the CERN machine with 14 GeV and luminosities of order 
300 fb$^{-1}$ unlike
the $q\bar q$ mode which will remain unobserved. While never competitive in size with the di-photon Higgs 
sample or the already known di-photon backgrounds, they may have to eventually be accounted for in high
precision measurements, particularly because the spin and CP-properties reconstructed from the two photons
in our reference $gg$-induced process are very similar to those of the Higgs signal, thus differently from
the case of the other backgrounds. Finally, we have shown the sensitivity of this channel to the certain
presence of the top quark and the possibile one
of additional heavy vector-like quarks (as an illustrative example) entering in the loops. 

In the longer term, the new channels studied here may also become quite significant in size, for example, after 
a few years of running at the so-called Super-LHC, a tenfold increase in istantaneous luminosity of the standard
14 TeV LHC \cite{SLHC}, which is currently being considered.

In summary, we have performed this calculation for the mere purpose of quantifying all possible sources of di-photon
events from the SM, no matter how small they could be, especially in presence of unsettled di-photon data  measurements. After all, recall
that the 125 GeV Higgs discovery was claimed on the basis of very few events, many from $\gamma\gamma$, 
so one would want that another Higgs boson is erroneously `discovered' 
in the di-photon channel
with apparent mass at 350 GeV (or elsewhere)
in a few years from now.  

\section*{Acknowledgments}
SM is supported in part through the NExT Institute. He is grateful to D.A. Ross for double-checking the calculation,
providing analytical formulae and innumerable useful comments on the analysis. We also acknowledge useful discussions with E. Accomando.


\newpage
\section*{Appendix: calculation of the integrals $C_0$ and $B_0$}

We start with
\beq
C_0(s,m^2)  \ \equiv \  -i \int \frac{d^4k}{\pi^2} \frac{1}{(k^2-m^2)((k-p_1)^2-m^2)((k+p_2)^2-m^2)},  
\eeq
where $p_1^2=p_2^2=0$ and $2p_1\cdot p_2=s$.

Upon Feynman parameterisation and by shifting the loop momentum to $k+p_1\alpha-p_2\beta$, one has

\beq C_0 \ = \  -2i \int \frac{d^4k}{\pi^2} \int_0^1 d\alpha d\beta \theta(1-\alpha-\beta)
 \frac{1}{\left[k^2+s\alpha\beta-m^2\right]^3 }.\eeq
Performing the loop integral, we get

\beq C_0 \ = \   \int_0^1 d\alpha d\beta \theta(1-\alpha-\beta)
 \frac{1}{\left[s\alpha\beta-m^2\right]}.\eeq
Integrating over $\beta$ gives  

\beq C_0 \ = \ \frac{1}{s} \int_0^1 \frac{d\alpha}{\alpha}
\ln\left(1-\frac{s\alpha(1-\alpha)}{m^2} \right).
\eeq

Defining 
   \beq\tau \ \equiv \ \frac{4m^2}{s} \eeq
 gives
  \beq C_0 \ = \ \frac{1}{s} \int_0^1 \frac{d\alpha}{\alpha}
\ln\left(1-\frac{4\alpha(1-\alpha)}{\tau} \right). \eeq

Factorising the argument of the logarithm produces 
  \beq C_0 \ = \ \frac{1}{s} \int_0^1 \frac{d\alpha}{\alpha} \left[
\ln\left(1-\frac{2\alpha}{1+\sqrt{1-\tau}} \right)
+\ln\left(1-\frac{2\alpha}{1-\sqrt{1-\tau}} \right)
   \right] \eeq
where we assume here that $\tau \, < \, 1$ so that we are above threshold
(i.e., we will have an imaginary part).

Then change variables to
 \beq \beta_1 \ = \ \frac{2\alpha}{1+\sqrt{1-\tau}} \eeq
for the first term and
 \beq \beta_2 \ = \ \frac{2\alpha}{1-\sqrt{1-\tau}} \eeq
for the second term to get
  \begin{eqnarray} C_0 & = & 
\frac{1}{s}  \left[ \int_0^{2/(1+\sqrt{1-\tau})} \frac{d\beta_1 \ln(1-\beta_1)}{\beta_1}
 + \int_0^{2/(1-\sqrt{1-\tau})} \frac{d\beta_2 \ln(1-\beta_2)}{\beta_2}
   \right]  \nonumber \\ &=&
     \frac{1}{s}  \left[
-\frac{\pi^2}{3} +\int_1^{2/(1+\sqrt{1-\tau})} \frac{d\beta_1 \ln(1-\beta_1)}{\beta_1}
 +  \int_1^{2/(1-\sqrt{1-\tau})} \frac{d\beta_2 \ln(1-\beta_2)}{\beta_2} \right].
 \nonumber \end{eqnarray}
Now change again variables to 
 $ \gamma_i \ \equiv \ \frac{1}{\beta_i} $ ($i=1,2$) 
to get
  \begin{eqnarray}  C_0 & =  & 
\frac{1}{s}  \left[ -
\frac{\pi^2}{3}
- \int_1^{(1+\sqrt{1-\tau})/2} \frac{d\gamma_1}{\gamma_1}  \ln\left(1-\frac{1}{\gamma_1}\right)
 - \int_1^{(1-\sqrt{1-\tau})/2} \frac{d\gamma_2}{\gamma_2}  \ln\left(1-\frac{1}{\gamma_2}\right)
 \right]
 \nonumber \\ &=&
\frac{1}{s}  \left[ -
\frac{\pi^2}{3}
- \int_1^{(1+\sqrt{1-\tau})/2} \frac{d\gamma_1}{\gamma_1} 
 \left( \ln\left(1-\gamma_1)\right) - \ln(-\gamma_1) \right)
- \int_1^{(1-\sqrt{1-\tau})/2} \frac{d\gamma_2}{\gamma_2} 
 \left( \ln\left(1-\gamma_2)\right) - \ln(-\gamma_2) \right) \right]
 \nonumber \\ &=&
\frac{1}{s}  \left[ -
\frac{\pi^2}{3}
 +\frac{1}{2} \ln^2\left(\frac{1+\sqrt{1-\tau}}{2}\right)
-i\pi \left| \ln\left(\frac{1+\sqrt{1-\tau}}{2}\right)\right|
 \right. \nonumber \\ & & \ \left.
 +\frac{1}{2} \ln^2\left(\frac{1-\sqrt{1-\tau}}{2}\right)
-i\pi\left|  \ln\left(\frac{1-\sqrt{1-\tau}}{2}\right) \right|
\right. \nonumber \\ & & \ \left.
- \int_1^{(1+\sqrt{1-\tau})/2} \frac{d\gamma_1}{\gamma_1} 
  \ln\left(1-\gamma_1\right)
- \int_1^{(1-\sqrt{1-\tau})/2} \frac{d\gamma_2}{\gamma_2} 
  \ln\left(1-\gamma_2\right)  \right] \label{bling},  \end{eqnarray}
where the sign of the imaginary part has been set to be negative
as required by unitarity.

Now we integrate the final term in (\ref{bling}) by parts 
 and rename the variable of integration $\gamma_2 \, \to \, 1-\gamma_1$
to achieve

  \begin{eqnarray}  C_0 & =  & 
\frac{1}{s}  \left[ -
\frac{\pi^2}{3}
 +\frac{1}{2} \ln^2\left(\frac{1+\sqrt{1-\tau}}{2}\right)\pm i\pi \ln\left(\frac{1+\sqrt{1-\tau}}{2}\right)
 \right. \nonumber \\ & & \ \left.
 +\frac{1}{2} \ln^2\left(\frac{1-\sqrt{1-\tau}}{2}\right)\pm i\pi \ln\left(\frac{1-\sqrt{1-\tau}}{2}\right)
\right. \nonumber \\ 
& & \ \left.
- \int_1^{(1+\sqrt{1-\tau})/2} \frac{d\gamma_1}{\gamma_1} 
  \ln\left(1-\gamma_1\right)
 + \int_0^{(1+\sqrt{1+\tau})/2} \frac{d\gamma_1}{\gamma_1} 
   \ln\left(1-\gamma_1\right) \right. \nonumber \\ & & \ \left.  
   -\ln\left(\frac{1+\sqrt{1-\tau}}{2}\right)
    \ln\left(\frac{1-\sqrt{1-\tau}}{2}\right) 
  \right] \nonumber \\ &  = & \ 
\frac{1}{s}  \left[ -
\frac{\pi^2}{2}
 +\frac{1}{2} \ln^2\left(\frac{1+\sqrt{1-\tau}}{2}\right)\pm 
i\pi \ln\left(\frac{1+\sqrt{1-\tau}}{2}\right)
 \right. \nonumber \\ & & \ \left.
 +\frac{1}{2} \ln^2\left(\frac{1-\sqrt{1-\tau}}{2}\right)\pm
 i\pi \ln\left(\frac{1-\sqrt{1-\tau}}{2}\right)
\right. \nonumber \\ & & \ \left.
 - \int_1^{(1+\sqrt{1-\tau})/2} \frac{d\gamma_1}{\gamma_1} 
  \ln\left(1-\gamma_1\right)
  + \int_1^{(1+\sqrt{1+\tau})/2} \frac{d\gamma_1}{\gamma_1} 
  \ln\left(1-\gamma_1\right)   \right. \nonumber \\ & & \   \left.
  -\ln\left(\frac{1+\sqrt{1-\tau}}{2}\right) 
   \ln\left(\frac{1-\sqrt{1-\tau}}{2}\right)
 \right] 
    \label{blong}.  \end{eqnarray}

We see that the remaining integrals (generating Spence functions) cancel.

Now use
\beq \left(\ln(-a)\right)^2+\left(\ln(-b)\right)^2-2\ln(a)\ln(b)
 \ = \ \left(\ln\left(\frac{a}{b}\right)\right)^2
 -2\pi^2 - i \pi \ln\left(\frac{a}{b}\right) \eeq
where the branch of the logarithm for negative argument is chosen to give a negative
imaginary part (again, the amplitude then has a positive imaginary part as required by
unitarity).

Finally, we have
\beq 
C_0 \  =  \ \frac{1}{2s}  \left[ 
  \ln\left(\frac{1+\sqrt{1-\tau}}{1-\sqrt{1-\tau}}\right)-i\pi \right]^2.
\eeq
Below threshold where $\tau \, > \, 1$ we make the substitution
\beq\ln\left(\frac{1+\sqrt{1-\tau}}{1-\sqrt{1-\tau}}\right)-i\pi
 \ \to  \ 2i \tan^{-1} \left(\frac{1}{\sqrt{\tau-1}} \right) 
 \ = \ 2i \sin^{-1} \left( \frac{1}{\sqrt{\tau}}\right).
\eeq

Similarly, start from 
\beq B_0(s,m^2) \ \equiv \frac{d^n k}{\pi^{n/2}} \frac{1}{(k^2-m^2)((k+p_1+p_2)^2-m^2)}
 + \frac{1}{(n-4)} (m^2)^{-(n-4)}. \eeq
Introducing Feynman parameters and shifting $k$ to $k+\alpha(p_1+p_2)$, one gets

\beq B_0(s,m^2) \ = \frac{d^n k}{pi^{n/2}}  \int_0^1 d\alpha
\frac{1}{\left[ (k^2-m^2+s\alpha(1-\alpha)\right]^2}
 + \frac{1}{(n-4)} (m^2)^{-(n-4)}. \eeq
Performing the loop integral  and cancelling the poles from the two terms
gives
\beq
 B_0 \ = \ -\int_0^1 d\alpha \ln\left(1-\frac{s\alpha(1-\alpha)}{m^2}\right). 
\eeq
Again, setting $\tau=4m^2/s$ and factorising the argument of the logarithm, we have

\begin{eqnarray} B_0 & = & -\int_0^1 d\alpha  \left[
\ln\left(1-\frac{2\alpha}{1+\sqrt{1-\tau}}\right) 
+\ln\left(1-\frac{2\alpha}{1-\sqrt{1-\tau}}\right) 
\right]. \nonumber \end{eqnarray}
Integrating over $\alpha$
\begin{eqnarray}  B_0
 &=&
2
-\frac{1+\sqrt{1-\tau}}{2} \left(1-\frac{2}{1+\sqrt{1-\tau}} \right) 
 \ln\left(1\frac{2}{1+\sqrt{1-\tau}} \right) 
 \nonumber \\ & & 
-\frac{1-\sqrt{1-\tau}}{2} \left(1-\frac{2}{1-\sqrt{1-\tau}} \right) 
 \ln\left(1\frac{2}{1-\sqrt{1-\tau}} \right) 
 \nonumber \\ & = & 
 2 - \sqrt{1-\tau}\ln\left(\frac{(1+\sqrt{1-\tau})}{(\sqrt{1-\tau}-1)} \right).
\nonumber  \end{eqnarray}
Again if $\tau \, > \, 1$ we make the replacement

\beq
\ln\left(\frac{(1+\sqrt{1-\tau})}{(\sqrt{1-\tau}-1)} \right)
 \ \to \ 2i \sin^{-1} \left(\frac{1}{\sqrt{\tau}} \right). 
\eeq

The expressions obtained for the $B_0$ and $C_0$ scalar integrals correspond to well known expressions used
in the case of a pseudoscalar Higgs boson entering $gg\to A\to \gamma\gamma$: practitioners would have recognised them.

\end{document}